# Modelling social-ecological transformations: an adaptive network proposal

Steven J. Lade[1,2], Örjan Bodin[1], Jonathan F. Donges[1,3], Elin Enfors Kautsky[1], Diego Galafassi[1], Per Olsson[1], Maja Schlüter[1]

1. Stockholm Resilience Centre, Stockholm University, Sweden
2. Fenner School of Environment and Society, The Australian National University, Australia
3. Potsdam Institute for Climate Impact Research, Potsdam, Germany

**Corresponding author:** Steven Lade, steven.lade@su.se, +46 70 191 9523

**Abstract**

Transformations to create more sustainable social-ecological systems are urgently needed. Structural change is a feature of transformations of social-ecological systems that is of critical importance but is little understood. Here, we propose a framework for conceptualising and modelling sustainability transformations based on adaptive networks. Adaptive networks focus attention on the interplay between the structure of a social-ecological system and the dynamics of individual entities. Adaptive networks could progress transformations research by: 1) focusing research on changes in structure; 2) providing a conceptual framework that clarifies the temporal dynamics of social-ecological transformations compared to the most commonly used heuristic in resilience studies, the ball-and-cup diagram; 3) providing quantitative modelling tools in an area of study dominated by qualitative methods. We illustrate the potential application of adaptive networks to social-ecological transformations using a case study of illegal fishing in the Southern Ocean and a theoretical model of socially networked resource users.

INTRODUCTION

In an era of rapid change in our planet's biophysical and socioeconomic systems (Steffen et al. 2015b), the ability of the Earth to sustain humanity is increasingly coming under threat (Steffen et al. 2015a). In response to these threats, large-scale transformations towards sustainability are needed within a range of fields and sectors (Schellnhuber et al. 2011), such as agriculture, fisheries, energy, and urban development. In this article we propose that adaptive networks can help analyse and understand sustainability transformations, in particular their structural dynamics: changes in the structure of the system over time.

Across the resilience, transitions, and pathways research literatures on sustainability transformations, three key aspects are generally mentioned (Fig. 1). First, *agency*: transformations do not 'just happen'; they need to be navigated and require active choices, skills and strategies from individuals and groups within the social-ecological system that is to be transformed. Much transformation literature is situated in discussions around agency, for example the literature on leadership, change agents and institutional entrepreneurship (Olsson et al. 2006; Westley et al. 2013). Second, *dynamics*: transformations involve changes over time (Grin et al. 2010; Leach et al. 2010). Historically, social-ecological dynamics have often been represented in the resilience literature using the ball and cup diagram to illustrate how the system might exist in different "regimes" or "stable states" (Walker et al. 2004). All three literatures successfully integrate agency and dynamics. For example, in the resilience literature, social-ecological transformations involve moving from one state to another, often through multiple phases, with individuals or groups strategically navigating the opportunity context in order to transgress certain thresholds (Olsson et al. 2006; Folke et al. 2009). Here, we focus on a resilience perspective on sustainability transformations, using this term and social-ecological transformation interchangeably.

The third key aspect of social-ecological transformations is *structure*: the pattern of interactions between the social and ecological entities that constitute a social-ecological system (Folke 2006). Sustainability transformations are widely understood to result in large-scale structural change (Folke et al. 2010). Moore et al. (2014) define transformation as "a form of change that … recombines existing elements of a system in fundamentally novel ways." For example, a study of a governance transformation in southern Sweden described how novel networking patterns between actors fostered the development of a new wetland governance regime (Olsson et al. 2004). As well as a result of transformation, structural change is also a key process during transformations. A study of the Southern Ocean by Österblom and Folke (2013) identified that changing patterns of interaction among a range of governmental and non-governmental actors led to radically different social-ecological outcomes in the context of overfishing.

A network approach is well suited to studying questions of structure. Networks provide a flexible framework, used in a variety of fields from sociology to neurosciences and medicine (Barabási et al. 2011; Prell 2012), which have already been used to study the connections between components of a social-ecological system (Lansing and Kremer 1993; Janssen et al. 2006; Bodin and Crona 2009; Cumming et al. 2010; Bodin and Tengö 2012), to analyse some structural aspects of sustainability transformations (Österblom and Folke 2013), and to analyse roles of agents in social-ecological networks (Bodin et al. 2014) . In these analyses, however, the networks were usually modelled statically, without explicit representation of the mechanisms that lead to changes in network structure over time. There remains a large gap in understanding the role of structural dynamics in sustainability transformations.

Here, we present a framework for transformations of social-ecological systems (and sustainability transformations more generally), based on adaptive networks, that can deal with structural dynamics. In adaptive network models, the structure (or topology) of the network and properties of nodes on the network co-evolve. Parallel literatures on adaptive networks have developed in the social network analysis literature (Snijders et al. 2010), where they are called dynamic networks or temporal networks, and more recently in the physics literature where they are called adaptive networks (Gross and Blasius 2008; Gross and Sayama 2009). The first theoretical adaptive network models of social-ecological systems (Wiedermann et al. 2015) are only now emerging, and case-based application of adaptive networks has not yet occurred.

Our aims in introducing an adaptive network framework for transformations are threefold. First, as discussed above, there is a need to integrate network structure, and especially structural dynamics, in our understandings of social-ecological transformations. Second, an adaptive network framework enables better communication of social-ecological change, particularly during transformations, as it is understood by resilience thinking. The most commonly used heuristic within resilience studies is the ball and cup diagram (Scheffer et al. 2001). While this has been a very widely used metaphor, which has successfully communicated aspects of persistence, adaptability and transformation, it has only to a limited degree been able to communicate that stability landscapes (the 'cups') themselves change over time, and the heuristic might therefore be interpreted as overly static in terms of existing system "states".

Third, while there have been rapid advances in qualitative analyses of transformations to sustainability, formal quantitative methods to study structural change in social-ecological systems are urgently needed. In the early development of resilience thinking, dynamical systems perspectives provided a rigorous theoretical basis for the ball and cup diagram (Gunderson 2000). At the same time, dynamical systems also provided a means of operationalising the ball and cup conceptualisation, allowing models to be developed and mechanisms explored. We anticipate that adaptive networks can play a similar role in the development of research on the social-ecological systems aspects of transformations to sustainability (Olsson et al. 2014). Furthermore, there are existing methods for adaptive network analysis that are well suited for immediate application to research on such systems. While methods such as agent-based modelling can incorporate both structure and dynamics, adaptive network approaches bring a set of statistical, analytical and conceptual tools that facilitate studying the interaction of dynamics and structure.

# ADAPTIVE NETWORK FRAMEWORK FOR TRANSFORMATIONS OF SOCIAL-ECOLOGICAL SYSTEMS

The structure and context of a social-ecological system are far from static, especially during a transformation. Trajectory illustrations have often been proposed to communicate the changing conditions in social-ecological systems over time (Leach et al. 2010), but they do not communicate underlying structural changes that shape social-ecological system behaviour. Frameworks that can help communicate and analyse structural dynamics are currently lacking.

For example, within resilience thinking, the ball and the cup diagram has become a widely used heuristic (Walker et al. 2004). In this diagram a ball, representing the social-ecological system moves about in a static 'landscape', consisting of a number of cups, representing the possible regimes in which the system could exist. The landscape is a result of the constraints placed on the system by its internal structure and dynamics, in interaction with its external context. In a classic ecological example, the state of a shallow lake (the 'ball') can be triggered into self-sustaining shifts between clear and turbid regimes (the 'cups') by nutrient input or by changes in species population sizes (Scheffer et al. 2001). While some authors have investigated situations in which the ball and cup diagram gradually changes (Scheffer et al. 2001), most applications of this diagram lack explicit acknowledgement of a temporal dimension, which is an essential element for understanding transformations. The lack of a temporal dimension may contribute to misguided interpretations of resilience thinking as a static perspective on social-ecological dynamics.

We propose adaptive networks as a framework for understanding the role of structural dynamics in social-ecological change, in particular transformations. Adaptive networks make structural dynamics explicit. In an adaptive network, changes in the social-ecological system's structure change the landscape in which the dynamics of the properties of individual components ('nodes') play out (hereafter, 'node dynamics'). Change in the shape of the landscape over time is a key advance over the static-landscape picture of resilience. Node dynamics on these landscapes, in turn, affect the future evolution of the system's structure (Fig. 2). Consider, for example, a social-ecological system in which a community of farmers cultivate individual plots of land (Wiedermann et al. 2015). The farmers interact, for example by trading goods or sharing cultivation practices (blue linkages in Fig. 2). Their farms are also connected biologically, for example by exchange of seeds, pollinators, and pests (green linkages in Fig. 2). Social and ecological connectivity affect the farming practices of individual farmers as well as the productivity of their land. At the same time, the structure of the network of farmers that trade or communicate with each other may be affected by changes of individual farmers' productivity or opinions regarding farming practices. This example is further developed below.

# TOWARDS ADAPTIVE NETWORK MODELS OF TRANSFORMATIONS

In addition to providing a new conceptualisation of social-ecological transformations, adaptive networks provide a tool with which quantitative dynamic models of transformations can be constructed. These models offer new opportunities to investigate how network structure and the properties of individual social and ecological entities co-evolve and participate in transformations. In this section, we describe how features of social-ecological systems could be implemented in an adaptive network model (Table 1), and briefly summarise the tools available to analyse these models.

**Social-ecological networks**

The foundation for an adaptive network model is a network conceptualisation of social-ecological systems. A core challenge for any network analysis is defining system boundaries and conceptualising what the nodes and links represent in the social-ecological system (Janssen et al. 2006; Bodin and Prell 2011; Bodin and Tengö 2012). In studies of social-ecological systems, 'social' nodes usually represent human actors or actor groups and 'ecological' nodes biophysical entities. In previous studies, social and ecological nodes have usually been each of a single type (such as clans and forest patches, respectively). Links usually represent interactions between these actors and entities, though they need not indicate a specific type of interaction but can indicate the likelihood of the nodes interacting, based for example on spatial proximity (Bodin and Tengö 2012).

In addition to the above examples, it should also be noted that the more 'static' notion of networks in social-ecological systems research has increased in interest over the last decade. For example, primarily with a social network focus, an increasing number of published work is concerned with analysing participation and collaboration in natural resource management/governance from an explicit network perspective (Bodin and Prell 2011).

**State and structure dynamics**

Adaptive network models extend a static network conceptualisation by introducing two network dynamics. First, the structure of the network can be dynamic, requiring rules for link re-wiring dynamics (the mechanisms that create, move or remove links between nodes; bottom row of Fig. 2) and possibly also rules for creation or removal of nodes. For example, a social network structure comprised by trade or communication between farmers could change depending on their productivity or opinions regarding farming practices. Ecological connectivity between farms could be altered by road construction. Second, the states of individual nodes can be dynamic, requiring conceptualising the states of nodes (which could be discrete, continuous, or categorical) and rules for the dynamics of node states (top row of Fig. 2). Changes in wellbeing of individual farmers, and the health of their farmland, can be affected by social (such as trade and communication networks) and ecological connectivity. Possible sources of dynamical rules for a social-ecological adaptive network are shown in Table 1.

**Existing methods**

Adaptive network models have been studied in two literatures: within the multidisciplinary social network literature in the social sciences, and the physics of complex systems literature. The research questions that adaptive networks are traditionally used to answer differ somewhat between these fields. Social network analysis typically asks, Given a series of observations of a network, such as a series of network snapshots, what link re-wiring rules are most likely to have created those dynamics (Snijders et al. 2010)? For example, what interactions between farmers are likely to have led to observed connectivity and productivity patterns? The physics literature, on the other hand, typically asks, Given a set of rules for link re-wiring and node state dynamics, what general properties are there of the resulting network dynamics (Gross and Sayama 2009)? For example, what patterns of network structure and farm productivity can emerge from social and ecological interactions between farms? A specific pattern, fragmentation, is explored in the stylised model introduced in the section 'Stylised model of adaptive network dynamics in a social-ecological system' below.

There exist well-established methods, corresponding to these research questions, which could be immediately applied to an adaptive network analysis of social-ecological transformation. Social network analysis emphasises statistical tests for link re-wiring rules (Snijders et al. 2010). The rules are represented using an objective function, which changes in network structure over time are assumed to optimise. This construction is well-suited to designing efficient statistical tests for network mechanisms. The physics literature emphasises mathematical techniques for producing general statements about network dynamics. Processes are typically defined independently, which is well-suited to introducing analytical techniques such as moment-closure or pair approximations (Demirel et al. 2014). Methods from ecology, where networks have been extensively studied, may also be useful (Rohr et al. 2014). We illustrate in the next section how these approaches to the analysis of adaptive networks could be applied to a social-ecological system.

**New methods: Network attractors**

There also exists substantial scope for further development of theoretical concepts and methods related to adaptive networks and their application to transformations in social-ecological systems. We briefly introduce two such examples here.

In a traditional dynamical systems interpretation of a ball-and-cup diagram, the attractors or 'cups' are assumed to correspond to attractors of the states of the social-ecological system (for example, resource levels, income levels, and so on). This is the sense in which we use ball and cup diagrams in Fig. 2. In a network attractor, the attractors would refer to persistent configurations or properties of network structure. For example, transitions between unsustainable and sustainable management could be characterised by different patterns of interactions between actors and the resource (Bodin and Tengö 2012) or between fragmented networks to more connected configurations (Olsson et al. 2007; Crona and Parker 2012). The notion of network attractors can help differentiating between adaptation and transformations. While adaptations imply changes in structures within the same network attractor (i.e. the general network configuration is maintained), transformations could be characterised as changes that move the system from one network attractor to the other. The agency of actors and their ability to establish new network attractors and navigate shifts to new system states seems to be crucial for transformations to sustainability (Westley et al. 2011).

A network attractor is analogous to a topological phase: qualitatively different configurations of networks that can change suddenly upon a small change of a parameter. Topological phase transitions have been investigated extensively in the physics literature on adaptive networks (Holme and Newman 2006; Bauke et al. 2011), for example fragmentation transitions in the context of coalition formation on dynamic networks (Auer et al. 2015; Schleussner et al. 2016). Properties of individual topological phases or network attractors, which may involve fixed-point or periodic dynamics, have also been investigated (Gross and Blasius 2008; Wiedermann et al. 2015). These studies, however, have generally been undertaken on large networks. In practice, however, social-ecological dynamics usually emerge from the interactions of a finite number of highly heterogeneous actors or entities. A network attractor concept for finite-size networks, which are arguably more relevant for social-ecological systems, has not been formally established.

**New methods: Distinguishing adaptation and transformation**

Alongside transformation, adaptation is another concept used by resilience thinking when describing structural change of social-ecological systems. Here, we focus on a structural understanding of adaptation. From a network perspective, adaptations are small incremental changes to the structure and functioning of a social-ecological system, while transformations correspond to large-scale reorganisation of the system structure that lead to a fundamental change in social-ecological feedbacks (Fig. 3) (Walker et al. 2004). Although increasingly addressed and explored by scholars of sustainability transformations, the distinction between adaptation and transformation is still vague. Some authors argue that adaptation and transformations are interrelated but operating at different scales (Folke et al. 2010). Although the distinction between adaptation and transformation is likely to always be to some degree dependent on context, question, and scale, we propose that changes in network measures such as connectivity could help to provide a functional and useful distinction between adaptation and transformation.

# APPLICATION TO MARINE GOVERNANCE IN THE SOUTHERN OCEAN

In this section, we discuss how adaptive network concepts and methods could be applied to the case study of illegal fishing in the Southern Ocean.

In the mid-1990s, illegal, unreported and unregulated (IUU) fishing in the Southern Ocean increased rapidly. As reported by Österblom and Folke (2013), non-governmental organisations (including environmental NGOs and the fishing industry) and state actors co-operated in response to this threat (Fig. 4). Interactions included collaborative monitoring, sharing operational information about specific vessels, co-ordinating strategies for stimulating political pressure, informal collaboration in criminal investigations and coordination of policy development (Österblom and Bodin 2012). A small number of actors started this collaboration informally, but actor networks increasingly became formalized in organisations focusing specifically on reducing illegal fishing, and their activities and capacities were increasingly integrated in the existing formal intergovernmental governance arrangement (Österblom and Folke 2013). These informal and formal governance networks increasingly developed their effectiveness (for example, by including new actors representing countries in which illegal activities had shifted to, or by including actors with necessary skills and competences) and were successful in reducing the illegal catch, as well as dealing with a series of resurgences in illegal activity (Österblom et al. 2010; Österblom and Sumaila 2011; Österblom and Folke 2013). The case has been described in terms of a governance transformation, radically improving the prospects for managing the fishery in a sustainable way.

In this system, social nodes could be the four types of governance actors (Fig. 4) and links would be the observed co-operation between them (Fig. 4). A range of link formation and re-wiring mechanisms could be explored (Table 1), for example based upon coalition formation. Depending on available data, some measure of willingness to take action against illegal fishing could be a relevant node state that is subject to influence from other nodes. The response of illegal fishers to governance actions would need to be operationalised in some form, if the cycles of governance action and later resurgence of illegal fishing are to be captured. Illegal fishers could for example constitute another node in the social-ecological network.

For the ecological part of the fishery, which was not investigated in detail by Österblom and Folke, relevant network components to be modelled could include: fish biomass, broken down by species or spatial location; trophic interactions or spatial exchange between these groups; and changes in the interactions between these groups (some examples are given in Table 1). Finally, a clearly important social-ecological link is illegal fishing itself.

Österblom and Folke (2013) focused on 'social' dynamics, presenting data on the governance network and the size of the illegal catch. Social network analysis methods, which are highly reliant on available data, could therefore be used for research questions related to the emergence of the new governance network, for example: What link formation mechanisms gave rise to the observed governance network dynamics? This analysis, using the statistical tools of social network analysis, would produce estimates on the likelihoods of particular rewiring mechanisms being present in this case, for example how coalitions form and dissolve. Research questions in the physics tradition, which focuses more on general behaviour and less on fitting a specific case, could include: Under which conditions does governance networks such as those in the Southern Ocean case emerge? Under what conditions are governance networks of the kind seen in the Southern Ocean case effective against illegal fishing? This model could have a variety of dynamical outcomes, including a consistently low illegal catch (as was observed), a consistently high illegal catch, or a perpetually fluctuating balance between illegal catch and governance activity.

# STYLISED MODEL OF ADAPTIVE NETWORK DYNAMICS IN A SOCIAL-ECOLOGICAL SYSTEM

The previous section speculated on how an adaptive network model of an empirical social-ecological system could be implemented. In this section, we describe a case in which an adaptive network model of a theoretical social-ecological system has been implemented. The model builds on the farming example introduced previously. The adaptive network clearly displays structural transformation as we have outlined above.

The COPAN:EXPLOIT model (Wiedermann et al. 2015; Barfuss et al. 2016) was specifically designed to study the emergent properties of a system of agents that each harvest a private renewable resource and, additionally, interact over an adaptive social network (Fig. 5A). This focus was motivated by asking how the dynamics of a system of only loosely or entirely unconnected resource users representing, for example, a preindustrial state of human societies, would change when social networks become increasingly dense and interactions increasingly fast as is reflected by the "Great Acceleration" leading to the Anthropocene (Steffen et al. 2015b).

In this model, each agent can either manage the logistically growing resource unsustainably, corresponding to short-term maximization of harvest rates, or manage sustainably leading to lower but long-term maintainable harvest rates. Agents interact with randomly chosen neighbours in the adaptive social network at a typical rate given by a prescribed social interaction time scale. Network structure and agent behaviour can change according to two important social processes, the balance of which is controlled by a social network rewiring parameter: (i) Homophily: when interacting with a neighbour with differing management preference, an agent can break the social tie and connect to another agent with identical preference. (ii) Boundedly rational imitation: agents can take over a neighbour's management preference, where the probability of imitation increases with the difference in current harvest rates, that is, agents imitate with high likelihood if the chosen neighbour currently harvests much more than themselves (the agents are myopic in this sense). See Wiedermann et al. (2015) for a detailed description of model structure, analytical approximations and results.

The model delivered the result, counter-intuitive at first glance, that faster social interactions increased the likelihood of ecological collapse (Fig. 5B). At fast social interaction rates, unsustainable harvest strategies, which initially return high payoffs, became dominant before the negative ecological effects of these strategies occurred. There was a sharp transition between globally sustainable and globally unsustainable outcomes. Crossing this tipping point led to transformative change in the social-ecological system, in the sense of a large and global change of accepted management preference. A fragmentation transition was also observed, where the social network decomposed into multiple disconnected groups of agents that each reached a local consensus state (Fig. 5B). In this way, the COPAN:EXPLOIT model illustrates that multiple types of transformative changes can be reflected already in a relatively simple adaptive network model of social-ecological dynamics. It furthermore demonstrates how social and ecological processes shape a desirable safe operating space in terms of sustainable management practices.

# DISCUSSION

In introducing an adaptive network framework for transformations of social-ecological systems, our aims were threefold. First, we sought to fill a gap in the analysis of transformations of social-ecological systems by introducing a framework that could conceptually integrate structural considerations with dynamics and agency and help analyse the interplay between them.

Second, we sought to counter unnecessarily static perceptions of resilience thinking, possibly perpetuated by the dominance of the ball and cup model of resilience. We identified a lack of conceptual models that can reflect resilience thinking on social-ecological transformations, a challenge that adaptive network models can meet. Adaptive networks also have the potential to make theoretical contributions, such as formally distinguishing between adaptation and transformation.

Third, we aimed to introduce quantitative modelling tools to an area of study that is currently dominated by qualitative analysis. Table 1 summarised some of components that can be used to develop an adaptive network model. While operationalising social-ecological systems in models remains a challenging process, the process of network building itself can be a useful process that integrates knowledge from various disciplines and fosters collaborative efforts to understand the dynamics of social-ecological transformations. While theoretical approaches can help systematically identify key factors and interactions influencing the emergent transformation, empirically-based studies such as the Southern Ocean example provided above can also provide empirical insight.

From a modelling perspective, adaptive networks are similar to agent-based models (ABMs) that are increasingly being used to study the dynamics of social-ecological systems (Schlüter et al. 2012; Schlüter et al. 2016). Agent-based models of human-environment systems generally consist of agents, their environment and interactions. Interactions between social agents can be represented as social networks that may change over time along with changes in agents' attributes. Social networks, for instance, can be important for human decision making as the actors within the network of one agent can have differential influence on its decisions (Matthews et al. 2007; An 2012). ABMs including dynamic social networks have been studied extensively in social simulation, particularly to understand the impact of network structure on opinion dynamics and the emergence of cooperation or social norms (Froncek 2015). ABMs of social-ecological systems, however, rarely focus explicitly on the role of networks as an emergent outcome or an explanatory factor for a particular social-ecological phenomenon. Exceptions include Caillault et al. (2013) on the effect of different networks on landscape patterns and Kaufmann et al. (2009) on the diffusion of land use practices. ABMs are often more complex and include multiple agent/node attributes and connections between them. While being very similar to agent-based models, adaptive networks specifically focus on structural dynamics and their consequences or explanatory power for system level dynamics such as transformations. Adaptive network models of social-ecological systems are in general less complex than ABMs which allows for mathematical analysis that cannot be applied to ABMs (Gross and Sayama 2009). This is an advantage with respect to the tractability of results, but comes at the costs of limitations in the complexity addressed. Adaptive networks can thus complement ABM by scrutinizing the role of structural changes, particularly of social or social-ecological networks, for large-scale social-ecological change such as transformations.

In summary, existing frameworks for social-ecological transformations have been useful, but fail to adequately bridge the dynamic and structural, aspects of transformations. We proposed an adaptive-network based framework that could bridge this gap. Adaptive networks could help analyse and develop further insights on how to move towards urgently needed sustainability transformations.

ACKNOWLEDGEMENTS

The research leading to these results received funding from: the European Research Council under the European Union's Seventh Framework Programme (FP/2007-2013)/ERC Grant Agreement 283950 SES-LINK; a core grant to the Stockholm Resilience Centre by Mistra; Project Grant 2014-589 from the Swedish Research Council Formas; the Stordalen Foundation (via the Planetary Boundary Research Network PB.net); and the Earth League's EarthDoc programme; the Strategic Research Programme EkoKlim at Stockholm University. Marc Wiedermann is acknowledged for his work on the COPAN:EXPLOIT model and for providing Fig. 5b; we thank Henrik Österblom for advice on the Southern Ocean case study.

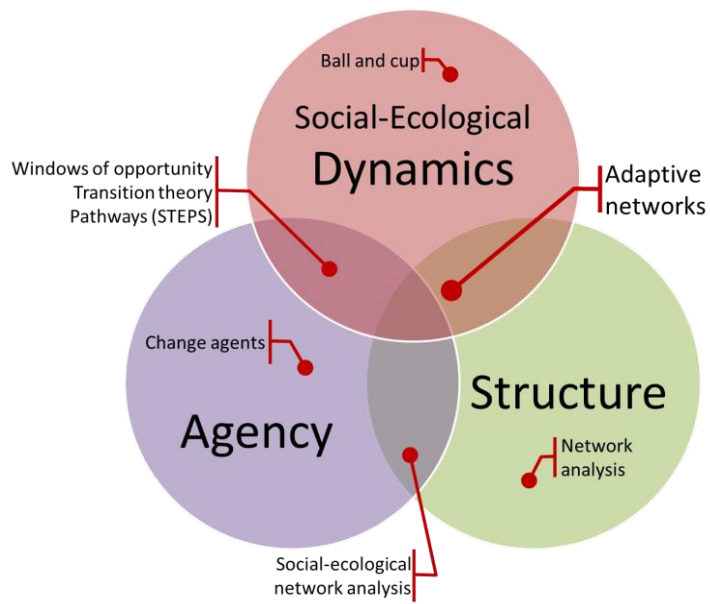

**Figure 1: What is a social-ecological transformation?** In this Venn diagram, we propose three key components of social-ecological transformation (coloured circles). Red dots (with callouts): Existing frameworks or concepts. In this article, we propose an adaptive network framework that is primarily designed to deal with structural dynamics, but to some degree can also integrate agency.

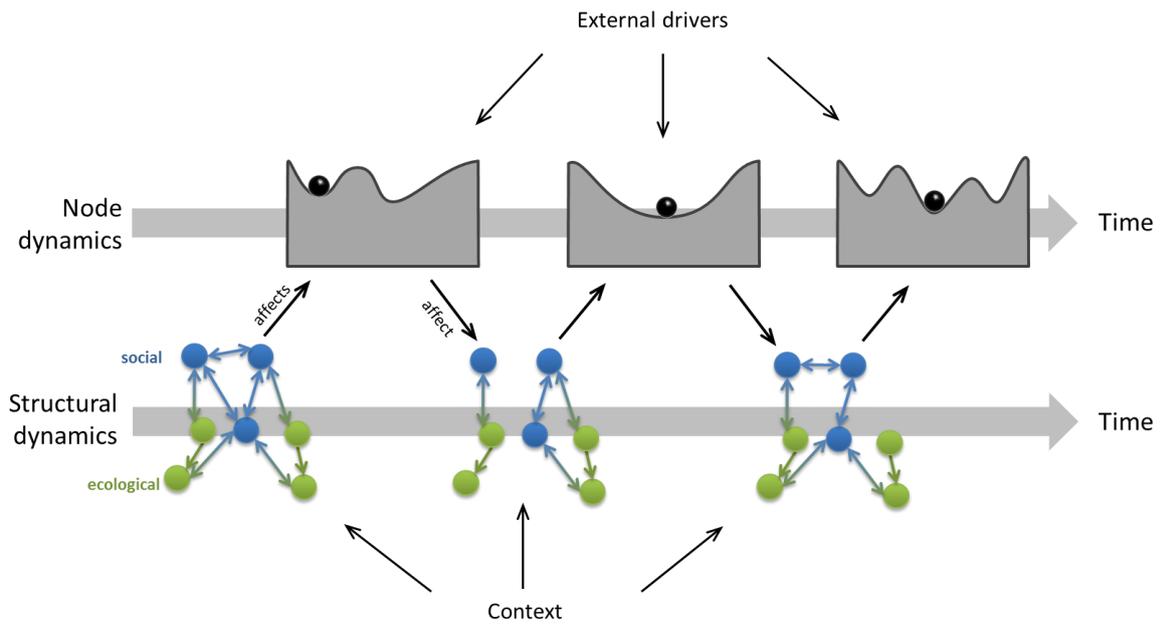

**Figure 2: Adaptive network framework.** In adaptive networks, there is a feedback between the dynamics of nodes on the network (node dynamics) and dynamics of the network structure (structural dynamics). The ball and cup diagrams represent possible dynamics of the properties of individual nodes of the network. While the framework emphasises internal dynamics, both node and structural dynamics can be influenced by external drivers as well as shaped by the context in which the system is situated.

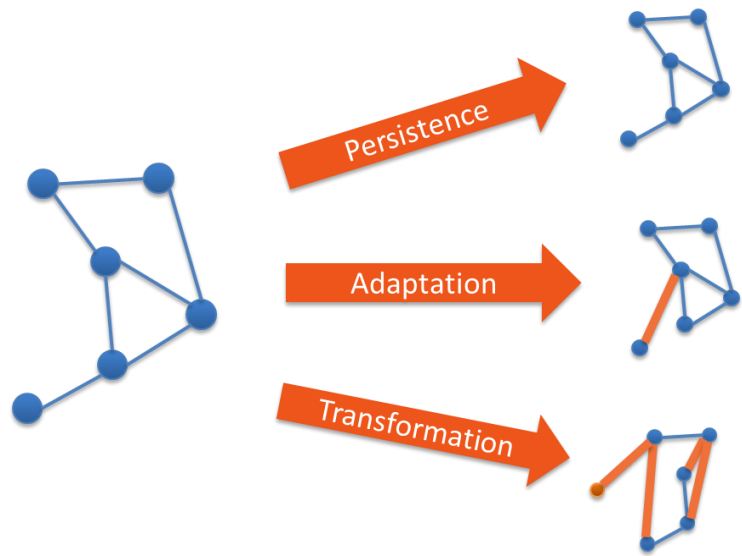

**Figure 3. Classification of social-ecological system dynamics.** An adaptive network approach could help distinguish between persistence, adaptation and transformation based on changes in network structure.

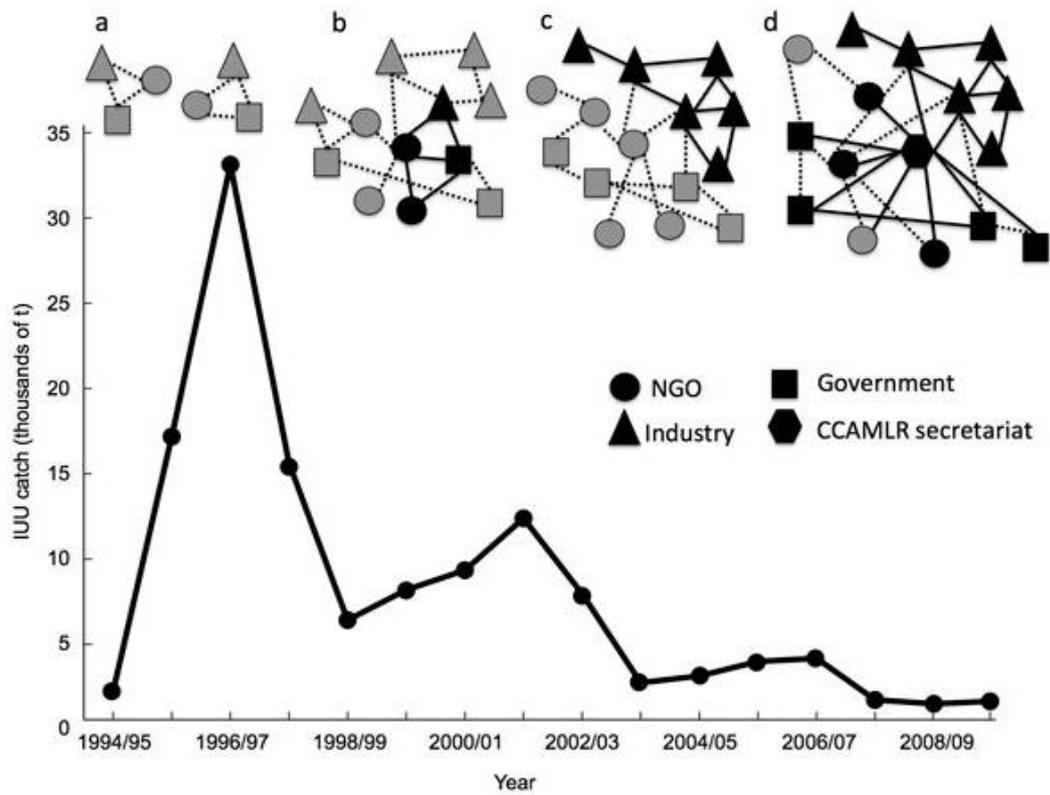

**Figure 4: Governance of fisheries in the Southern Ocean.** Illegal, unregulated and unreported (IUU) catch in the Southern Ocean between 1995 and 2009. Qualitative networks (a-d) represent the evolution of formal (black) and informal (grey) governance network involving state and non-state actors and organisations, which occurred at a = 1996, b =1998, c = 2003 and d = 2005. NGO = Non-Governmental Organization; CCALMR = Commission for the Conservation of Antarctic Marine Living Resources. Reproduced from Österblom and Folke (2013).

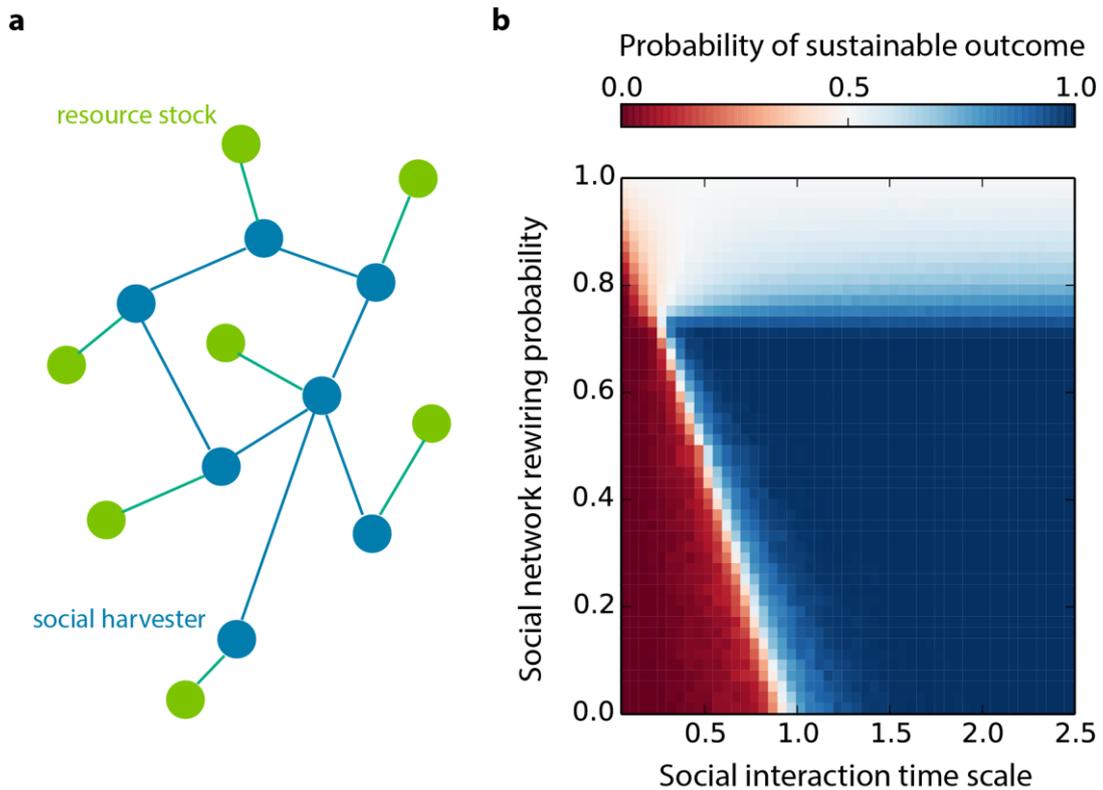

**Figure 5: The EXPLOIT model.** (a) General model structure: agents (social nodes, blue) harvesting renewable private resource stocks (ecological nodes, green) interact via an adaptive social network by boundedly rational imitation of management practise (sustainable or short-term maximization of harvest rates) and homophilic rewiring of social ties. The illustration shows one possible instance of the network structure, which changes over time. (b) Varying key model parameters such as social interaction time scale (measured in units of resource regeneration time scale) and social network rewiring probability can lead to transformative change of social-ecological system structure.

**Table 1: Components of an adaptive network model.** We describe the components of an adaptive network model with social and ecological (and, where appropriate, social-ecological) examples.

| Component | Examples |
|---|---|
| Nodes (or 'vertices') | Social: Individuals, communities or other groups, nations<br>Ecological: Individual animals, species, resource patches |
| Links (or 'edges') | Social: Trade, communication, social support<br>Ecological: Animal movement, species competition, trophic interactions, resource flow<br>Social-ecological: Resource extraction, pollution, ecosystem service utilization, observation, management |
| States of nodes | Social: Opinion, preferences, size of group, coalition membership<br>Ecological: Population size, biomass, biodiversity |
| Rules for link re-wiring dynamics | Social: Homophily, heterophily, transitivity, reciprocity, assortative matching, acquiring resources<br>Ecological: Adaptive prey switching, adaptive foraging, adaptive resource responses |
| Rules for node dynamics | Social: Opinion dynamics, innovation adoption, coalition formation<br>Ecological: Population dynamics, epidemic models, |
| Rules for node creation or removal | Social: Innovation of new technologies or ideas bringing new actors into the system, creation or destruction of organisations, in- and out migration<br>Ecological: Evolution, immigration, extinction |